\documentclass[12pt]{article}
\usepackage{epsfig,amsfonts,amssymb}
\usepackage{hyperref}
\pdfoutput=1

\usepackage{cite}
\usepackage{cite}

\usepackage{comment}

\def\ZZZ{{\hbox{ Z\kern-1.6mm Z}}}
\def\RRR{{\hbox{ R\kern-2.4mm R}}}
\def\CCC{{\hbox{ C\kern-2.0mm C}}}
\def\zzz{{\hbox{z\kern-1mm z}}}

\newcommand{\qeq}{{\hbox{=\kern-2.3mm ? \kern.5mm }}}
\renewcommand{\qeq}{=}

\newcommand{\DD}{{\cal D}}

\newcommand{\MM}{{\cal M}}

\newcommand{\NN}{{\cal N}}

\newcommand{\be}{\begin{equation}}
\newcommand{\ee}{\end{equation}}
\newcommand{\ben}{\begin{eqnarray}\displaystyle}
\newcommand{\een}{\end{eqnarray}}

\newcommand{\refb}[1]{(\ref{#1})}
\newcommand{\p}{\partial}
\newcommand{\sectiono}[1]{\section{#1}\setcounter{equation}{0}}

\def\one{{\hbox{ 1\kern-.8mm l}}}
\def\zero{{\hbox{ 0\kern-1.5mm 0}}}

\newcommand{\bea}[1]{\begin{eqnarray}\label{#1} }
\newcommand{\eea}{\end{eqnarray}}

\newcommand{\eqref}{\refb}

\newcommand{\q}{e}

\usepackage{bm}
\usepackage[table]{xcolor}

\begin{document}

\baselineskip 24pt

\begin{center}

{\Large \bf Revisiting Logarithmic Correction to Five Dimensional BPS Black Hole Entropy}

\end{center}

\vskip .6cm
\medskip

\vspace*{4.0ex}

\baselineskip=18pt

\centerline{\large \rm A.H.~Anupam,  Chandramouli Chowdhury  and Ashoke Sen}

\vspace*{4.0ex}

\centerline{\large \it International Centre for Theoretical Sciences - TIFR 
}
\centerline{\large \it  Bengaluru - 560089, India}

%\centerline{\large \it ~$^c$Homi Bhabha National Institute}
%\centerline{\large \it Training School Complex, Anushakti Nagar,
%    Mumbai 400085, India}

\vspace*{1.0ex}
\centerline{\small E-mail: anupam.ah,
chandramouli.chowdhury,
 ashoke.sen \ @icts.res.in}

\vspace*{5.0ex}

\centerline{\bf Abstract} \bigskip

We compute logarithmic correction to the entropy of  BPS black holes 
in asymptotically flat five dimensional space-time using
finite temperature black hole geometry and find perfect agreement with the microscopic
results and macroscopic computations based on zero temperature near horizon geometry.
We also reproduce the Bekenstein-Hawking term for zero temperature black hole entropy
from the corresponding term for finite temperature black hole.

\vfill \eject

\tableofcontents

\sectiono{Introduction}

Bekenstein-Hawking formula gives the leading contribution to the black hole entropy.
It is expected to receive corrections that are suppressed by inverse powers of the black
hole size and also corrections proportional to the logarithm of the black hole size.
Logarithmic correction to the black hole entropy has proved a fruitful direction of research in the
past since it can be computed purely from the infrared data of the theory -- massless fields
and their interaction. Yet it puts strong constraint on any ultra-violet completion of the theory
by requiring that explicit
counting of microstates must reproduce the same correction. This has been tested for
supersymmetric black holes for many four and five dimensional string
compactifications\cite{1005.3044,1106.0080,1109.3706,1404.1379,1402.2441,1404.6363}.

Traditionally the computation
of the logarithmic corrections to the entropy of
non-extremal and extremal (BPS) black holes have used
somewhat different routes, with the former employing gravitational path integral in the
full asymptotically flat space-time geometry and the latter using the near horizon geometry
of the black hole. This is necessitated by the fact that  
in the full geometry of extremal black holes
there are two large length scales -- the size of the
black hole and the inverse temperature that is taken to infinity, and it is technically challenging
to separate the logarithm of the size of the black hole from the logarithm of the 
temperature.\footnote{Some recent analysis of the logarithm of the temperature can be found
in \cite{Iliesiu:2022onk,Banerjee:2023quv}.}
Logarithmic corrections to BPS black hole entropy computed from the near horizon geometry
could be tested against the results of explicit counting of
microstates in supersymmetric string compactifications, 
while for non-supersymmetric black holes far from extremality, 
there is no independent microscopic computation
in string theory.

This situation changed with the work of \cite{2107.09062}
which proposed a way to
compute the index of supersymmetric states\footnote{Throughout this paper what we 
call index is degeneracy counted
with weight $\pm 1$ depending on the whether the state is a bosonic or fermionic supermultiplet.
This is the quantity computed in the microscopic counting and is protected against quantum
corrections. We shall also call  the logarithm of the index as entropy.} 
using gravitational path integral in the full asymptotically flat space-time geometry. 
(Earlier work along the same line  for AdS spaces can be found in  \cite{Cabo-Bizet:2018ehj} and for  near
horizon geometry can be found in \cite{Heydeman:2020hhw}).
The idea is
to work at a finite temperature and introduce a chemical potential conjugate to the
angular momentum that introduces a factor of $(-1)^F$ in the trace computed by the
gravitational path integral.  With some additional factor of angular momentum inserted
into the path integral (discussed in section \ref{sindex}) this is precisely the index that counts
supersymmetric ground states. Since this analysis can be
done at any temperature, we do not encounter the problem we had with extremal black
holes and in particular can choose the inverse temperature to be controlled by the same
length scale that controls the black hole size.

In a previous paper we carried out this analysis for four dimensional black holes in
asymptotically flat space-time
and found
perfect agreement between the near horizon calculation and the full geometry 
calculation of the logarithmic correction to the entropy\cite{H:2023qko}.
This also means that the full geometry calculation gives result in agreement with the result
of microscopic calculation whenever the latter results are available. In this paper we shall
use the solutions constructed in \cite{Cvetic:1996xz} to
extend the analysis to five dimensional black holes in asymptotically flat space-time,
and again find perfect agreement with the
earlier results based on the near horizon geometry of the black hole solutions
as well as microscopic counting as
described in \cite{1109.3706}. While in spirit
the analysis is similar to that in four dimensions, we find that in five dimensions the change of
ensemble produces non-vanishing logarithmic corrections and must be taken into account
in order to get agreement with the earlier results. This also confirms the general procedure
that was used earlier for computing logarithmic corrections to non-extremal black hole 
entropy. 

As part of our analysis we also check that the classical entropy of the extremal
black hole comes out correctly from the saddle point representing non-extremal black holes.
The agreement is somewhat non-trivial since the agreement is not between the entropies
computed on the two sides, but between the entropy of an extremal black hole and the
entropy of a non-extremal black hole plus a term proportional to the angular momentum
carried by the black hole\cite{2107.09062}.

The rest of the paper is organized as follows. In section \ref{sscale} we review the
scaling properties of various thermodynamic parameters of a black hole in arbitrary
dimension. This determines how we should take the macroscopic limit of the various
parameters so that the black hole metric scales by an overall constant factor in this limit.
In section \ref{sindex} we review the general formalism given in \cite{2107.09062} for
computing the supersymmetric index of the black hole from gravitational 
partition function. Particular emphasis is placed on the point that in order to go from the
partition function to the index, we need to undergo a change of ensemble, and this
produces some non-trivial logarithmic terms that must be taken into account.
In section \ref{sclass} we examine the classical limit of the relation
between index and partition function. This leads to a non-trivial relation between the
entropies of zero temperature and finite temperature black holes and we verify this
relation using the known five dimensional solutions of 
\cite{Cvetic:1996xz,Cvetic:1996kv}. In section \ref{slog} we compute the logarithmic correction to
the partition function for a class of supersymmetric black holes in five dimensions, and
combine this result with the result of section \ref{sindex} to compute the logarithmic correction
to the index of these black holes. The result agrees perfectly with the earlier results based
on the near horizon analysis and also the results of microscopic counting. In appendix \ref{sa}
we review the non-extremal five dimensional black hole solution of
\cite{Cvetic:1996xz,Cvetic:1996kv}, and compute various quantities that are needed to
test the relation between extremal and non-extremal black hole entropy, as discussed in
section \ref{sclass}.

\sectiono{Scaling of the black hole parameters} \label{sscale}

In $D$ dimensions the classical entropy $S_0$ of the black hole, given as a function of the
charges $\vec Q$, mass $M$ and angular momenta $\vec J$ in the Cartan subalgebra of the rotation
group,  scales as\cite{1205.0971}
\be\label{e1.1}
S_0(\lambda^{D-3}M, \lambda^{D-3}\vec Q, \lambda^{D-2}\vec J)
= \lambda^{D-2} S_0(M, \vec Q, \vec J)\, .
\ee
The vector symbol on $\vec J$ and $\vec\Omega$ 
labels different
components of the angular momentum and angular velocity 
inside the Cartan subalgebra. This should not be confused
with the vector symbol on $\vec J_L$, $\vec J_R$ in later sections where it will denote
all generators inside the
$SU(2)_L$ and $SU(2)_R$ subgroups of the five dimensional rotation group. Throughout
this paper we shall follow the convention that $\vec J$ without a subscript denotes all
generators of the Cartan subalgebra while $\vec J_L,\vec J_R$ denote all generators of
$SU(2)_L$ and $SU(2)_R$ respectively. The vector symbol on $\vec Q$ is a reminder of
the fact that the theory may have multiple $U(1)$ gauge fields and the black hole may 
be charged under more than one of these gauge fields.
$\vec Q$ is an $n_V$ dimensional vector if the
theory has $n_V$ U(1) gauge fields.

\refb{e1.1} suggests that 
it is useful to introduce new parameters $m$, $\vec q$, $\vec j$ via
\be\label{e1.2}
M = \lambda^{D-3} m, \qquad \vec Q =\lambda^{D-3}\vec q, \qquad \vec J =\lambda^{D-2}
\vec j\, ,
\ee
and define the macroscopic limit to be the large $\lambda$ limit keeping $m,\vec q, \vec j$
fixed at order one. Also in this limit the temperature $\beta$, chemical potential $\vec \mu$ and angular
velocity $\vec\Omega$ scale as
\be\label{e1.3}
\beta = {\p S_0\over \p M}\sim \lambda, \qquad \vec\mu = {1\over \beta}
{\p S_0\over \p \vec Q} \sim 1, \qquad \vec\Omega = {1\over \beta}
{\p S_0\over \p \vec J} \sim \lambda^{-1}\, .
\ee

BPS black holes have infinite
temperature and the mass and entropy are expressed as functions of the charges and the
angular momenta. All quantities other than the temperature 
satisfy the same scaling relations, e.g.
\be
\vec Q \sim \lambda^{D-3}, \qquad \vec J \sim \lambda^{D-2}, \qquad
M_{BPS}\sim \lambda^{D-3}, \qquad S_{BPS}\sim \lambda^{D-2}\, .
\ee

For our analysis we shall also need the scaling properties of the black hole
partition function, defined as
\be\label{edefz0}
\ln Z_0(\beta,\vec\mu,\vec\Omega) = S_0 -\beta M -\beta\vec \mu.\vec Q -\beta\vec\Omega.
\vec J\, .
\ee
It follows from the results given above that $\ln Z_0$ scales as
\be\label{ezscale}
\ln Z_0\sim \lambda^{D-2}\, .
\ee

\sectiono{Index from partition function} \label{sindex}

In five dimensions
the rotation group is $SO(4) = SU(2)_L\times SU(2)_R$ where the subscripts L and R 
have been introduced for convenience. A generic black hole in five dimensions carry three
$U(1)$ charges,  mass $M$ and two angular momenta $J_\phi$ and 
$J_\psi$
in orthogonal planes with azimuthal angles $\phi$ and $\psi$\cite{Cvetic:1996xz}. 
The third components of 
$SU(2)_L$ and $SU(2)_R$ angular momenta are
given in terms of $J_\phi$ and $J_\psi$ as
\be
J_{3L}={1\over 2} (J_\phi-J_\psi), \qquad J_{3R} = {1\over 2} (J_\phi+J_\psi)\, .
\ee
The conjugate variables are
the inverse temperature $\beta$, the chemical potential $\vec\mu$ for the charges $\vec Q$
and the chemical potentials (angular velocities) $\Omega_L,\Omega_R$ conjugate to $J_{3L},J_{3R}$. Classical supersymmetric
black hole solutions have $J_{3R}=0$.

The relevant index for a supersymmetric black hole that breaks $2n$ $SU(2)_L$
invariant supersymmetries  is\cite{Dabholkar:2010rm}:\footnote{We could have used the
more conventional definition $(-1)^F=e^{2\pi i (J_{3L}+J_{3R})}$ in \refb{edefnbps} since the
$e^{2\pi i J_{3L}}$ factor will just give an overall phase when we compute trace over states
with fixed $J_{3L}$. But in subsequent analysis we shall stick to the definition of $(-1)^F$
given in \refb{edefnbps}.}
\be\label{edefnbps}
N_{BPS}(\vec Q, J_{3L}) \equiv e^{S_{BPS}(\vec Q, J_{3L})} 
= Tr_{\vec Q, J_{3L}, \vec k=0} \left[(-1)^F (2 J_{3R})^n\right]\, , \qquad (-1)^F \equiv 
e^{2\pi i J_{3R}}\, ,
\ee
with the trace taken over all states carrying fixed $(\vec Q, J_{3L})$ and zero momentum.
This generalizes the helicity supertrace index in four dimensions\cite{9611205,9708062}. 
The sum is
expected to pick up contribution only from BPS states that break $2n$ or less $SU(2)_L$
invariant supersymmetries. Therefore $N_{BPS}=e^{S_{BPS}}$ counts
BPS states with a fixed charge vector $\vec Q$,
a fixed $J_{3L}$ and zero momentum, but all values of
$\vec J_{L}^2$, $J_{3R}$ and $\vec J_R^2$, weighted by $(-1)^F (2 J_{3R})^n$.
 We shall refer to this as the index.

For computing $N_{BPS}$ from the macroscopic side, 
we begin with the gravitational partition 
function $Z$ with asymptotic
boundary conditions appropriate to that of an Euclidean black hole
with temperature $\beta$, chemical potentials $\mu$ and angular velocities $\Omega_{L,R}$,
with
a factor of $(2J_{3R})^n$ inserted into the path integral. 
The relevant boundary condition requires 
the Euclidean time $\tau$ and the azimuthal angles $\vec\phi$
conjugate to $\vec J$  to be periodically identified as
\be 
(\tau,\vec\phi)\equiv (\tau+\beta, \vec \phi-i\beta\, \vec\Omega)\, ,
\ee 
and the $\tau$ component of the gauge field at infinity to be fixed at
\be
\vec A_\tau = -i\vec\mu\, .
\ee
$Z$ defined this way computes:
\be\label{eorig}
Z(\beta,\vec\mu, \Omega_L,\Omega_R)
 = Tr \left[e^{-\beta E - \beta\vec\mu.\vec Q -\beta \Omega_L J_{3L} - \beta \Omega_R J_{3R}} (2J_{3R})^n\right] \, ,
\ee
where the trace is taken over all the states. 
For computing the index $N_{BPS}=e^{S_{BPS}}$ we 
follow \cite{2107.09062}
and set the chemical potential $\Omega_R$ dual to $J_{3R}$ to $-2\pi i/\beta$. 
This computes
\ben\label{eorig1}
Z(\beta,\vec\mu, \Omega_L,-2\pi i/\beta)
 &=& Tr \left[e^{-\beta E - \beta\vec\mu.\vec Q -\beta \Omega_L J_{3L} + 2\pi i J_{3R}} (2J_{3R})^n\right] \nonumber \\
& =& Tr \left[e^{-\beta E - \beta\vec\mu.\vec Q -\beta \Omega_L J_{3L}} (-1)^F (2J_{3R})^n\right] \, .
\een
Even though the trace runs over all states of the theory,
it is expected to pick the contribution only from the BPS states. Let us denote by
$M_{BPS}(\vec Q, J_{3L})$  the mass of a BPS state carrying
quantum numbers $(\vec Q, J_{3L})$. We can organize the trace in the expression
for $Z$ by first taking the trace over all quantum numbers other than $(\vec Q, J_{3L},\vec k)$,
and then sum over $\vec Q, J_{3L}$ and integrate over $\vec k$. 
Since the first step produces $N_{BPS}(\vec Q, J_{3L})$,
this allows us to express $Z$ as a weighted sum over $N_{BPS}(\vec Q, J_{3L})
= e^{S_{BPS}(\vec Q, J_{3L})}$.
The $\vec k$ integral can be easily done by noting that the $\vec k$ dependence of the
integrand comes
only through the dependence of $E$ on $\vec k$.
We can replace $E$ by $M_{BPS}+ \vec k^2 / (2 M_{BPS})$ and
express \refb{eorig1} as,
\be\label{estra}
Z(\beta,\vec\mu, \Omega_L,-2\pi i/\beta) =\sum_{\vec Q, J_{3L}} 
e^{S_{BPS}-\beta M_{BPS}
-\beta \vec\mu.\vec Q-\beta\Omega_L J_{3L}} \int d^{n_T} k \left({L\over 2\pi}\right)^{n_T} e^{-\beta\vec k^2 / (2M_{BPS})} \, ,
\ee
where $\vec k$ is an $n_T$ dimensional momentum vector that is invariant the rotation
group element $e^{-\beta\Omega_L J_{3L}}$ 
and $L$ is the physical size of the box in which we 
place the black hole. This restriction on $\vec k$ stems from the fact that if $\vec k$ is
not invariant under $e^{-\beta\Omega_L J_{3L}}$, then the action of
this operator will produce a state with different $\vec k$ and hence such a state
will not contribute
to the trace. 
We shall see that the final result for logarithmic correction is independent of $n_T$.

After performing the integration over $\vec k$,  \refb{estra} can be inverted as 
\be\label{enexp}
e^{S_{BPS}} \sim L^{-n_T} \left({\beta\over M_{BPS}}\right)^{n_T/2}
\beta^{n_V+1} \int d^{n_V} \mu\, d\Omega_L \, 
e^{\beta M_{BPS}
+\beta \vec\mu.\vec Q+\beta\Omega_L J_{3L}
+ \ln Z(\beta,\vec\mu, \Omega_L,-2\pi i/\beta)} \, ,
\ee
with appropriate choice of integration  contours for $\vec \mu$ and $\Omega_L$. We 
have ignored overall numerical factors since they contribute constant terms in $S_{BPS}$ and
are not of interest for the current paper.
 \refb{estra} and \refb{enexp} 
  holds for all $\beta$ but we shall take $\beta$ to scale as $\lambda$ as in
 the case of non-extremal black holes given in \refb{e1.3} to facilitate our analysis.

The leading classical result for $\ln Z$ is given by $\ln Z_0$ given in \refb{edefz0}. 
If we replace $Z$ by $Z_0$ in \refb{enexp} and carry out the integration over 
$\vec k$, $\vec \mu$ and $\Omega_L$ using saddle point approximation, we get
\ben\label{eintegral}
e^{S_{BPS}} &\sim&  L^{-n_T} \left({\beta\over M_{BPS}}\right)^{n_T/2} \beta^{n_V+1}
\left(\det { \p^2 \ln Z_0\over \p \mu_i \p\mu_j}\right)^{-1/2}
 \, \left( {\p^2 \ln Z_0\over \p \Omega_L^2}\right)^{-1/2} \nonumber\\ &\times&
 e^{\beta M_{BPS}
+\beta \vec\mu.\vec Q+\beta\Omega_L J_{3L}
+ \ln Z_0(\beta,\vec\mu, \Omega_L,-2\pi i/\beta)}\, ,
 \een
 evaluated at the saddle point
 \be\label{esaddle}
{1\over \beta}  {\p \ln Z_0\over \p \mu_i}=-Q_i, \qquad  {1\over \beta} 
{\p \ln Z_0\over \p \Omega_L}=- J_{3L}\, .
 \ee
 Due to \refb{edefz0} these equations are the same as the ones in \refb{e1.3}. Using
 the various scaling properties described in section \ref{sscale}, and setting $D=5$, we get
 \ben
 S_{BPS} &\simeq&  \beta M_{BPS}
+\beta \vec\mu.\vec Q+\beta\Omega_L J_{3L}
+ \ln Z_0  \nonumber\\ && \hskip .5in -n_T \ln L -
{n_T\over 2} \ln\lambda - {n_V\over 2}\ln\lambda -{3\over 2} \ln\lambda + \delta \ln Z\, ,
\een
where $\delta\ln Z$ denotes further logarithmic corrections from integration over
massless fields in
the gravitational path integral and inclusion of the $(2J_3)^n$ factor in the definition of $Z$.
Using \refb{edefz0} and 
$\beta\Omega_R=-2\pi i$,
we get the sum of the leading contribution
and logarithmic contribution to $S_{BPS}$ in $D=5$ as\footnote{In the presence of such
corrections, \refb{esaddle} also gets modified, but these can be shown not to affect the logarithmic
correction to the entropy.}
 \ben\label{ebpsfin}
 S_{BPS} &\simeq&  \beta M_{BPS} + S_0 - \beta\, M + 2\pi i J_{3R} 
 \nonumber\\ && -n_T \ln L-
{n_T\over 2} \ln\lambda - {n_V\over 2}\ln\lambda -{3\over 2} \ln\lambda
+\delta \ln Z\, .
\een
Note that $M$ and $S_0$ are the entropy of a classical black hole carrying temperature
$\beta$, $\beta\Omega_R=-2\pi i$, charges $\vec Q$ and $SU(2)_L$ charge $J_{3L}$,
and are not {\it a priori} the same as $M_{BPS}$ and $S_{BPS}$ which are properties
of a zero temperature black hole.

It should be understood that the equality in \refb{ebpsfin} 
holds for the Bekenstein-Hawking term and the
term proportional to $\ln\lambda$. In particular in five dimensions we also expect terms
linear in $\lambda$ that will not be kept track of. Even though these terms dominate over
$\ln\lambda$ for large $\lambda$, they are sensitive to the details of the UV completion, {\it e.g.}
higher derivative terms in the theory, and are not computable just from the low energy data.
We can get rid of such polynomial terms by taking sufficient number of derivatives with respect
to $\lambda$, so that at the end the dominant term comes from the coefficient of the $\ln\lambda$
term.

It is useful to compare \refb{ebpsfin} with similar result in $D=4$. For this we can examine the 
result of the integrals in \refb{eintegral} for general $D$, and modify \refb{ebpsfin}
accordingly. This would give, 
 \ben\label{eloggen}
 S_{BPS} &\simeq&  \beta M_{BPS} + S_0 - \beta\, M + 2\pi i J_0
+\delta \ln Z  \nonumber\\ &&  \hskip -.5in -n_T(D-4) \ln L -
{n_T\over 2} (D-4) \ln\lambda - {n_V\over 2} (D-4)\ln\lambda -{1\over 2}(D-2)(n_C-1) \ln\lambda\, .
\een
Here $J_0$ stands for whatever plays the role of $J_{3R}$ in a general dimension, {\it e.g.} in four
dimensions where the rotation group is $SU(2)$, we take $J_0$ to be $J_3$.
$n_C$ is the dimension of the Cartan subalgebra of the rotation group, and the $(n_C-1)$
factor gives
the dimension of the analog of $\Omega_L$ in general dimension. In $D=4$ we have $n_C=1$
and hence all the logarithmic corrections in the second line of
\refb{eloggen} vanish, leaving behind $\delta \ln Z$ as the only source of logarithmic
corrections. Therefore the logarithmic terms coming from change of ensemble, as discussed
in this section, are absent
in four dimensions.

\sectiono{Classical result} \label{sclass}

In this section we shall explore the consequences of the classical result $S^{(0)}_{BPS}$
for $S_{BPS}$
given by the first
line of \refb{ebpsfin}:
\be\label{enfin}
S^{(0)}_{BPS} = S_0 - \beta M +\beta M_{BPS}
+2\pi i J_{3R}\, .
\ee
In this equation 
$M$ denotes the mass of the black hole solution corresponding to the saddle point
that contributes to the index. On physical grounds we expect this to be equal to the BPS
mass $M_{BPS}$:
\be\label{efirst}
M = M_{BPS}\, .
\ee
However, this 
equality is in no way obvious since {\it a priori} $M$ depends on $\beta$ which can be
chosen arbitrarily.
If \refb{efirst} holds then 
\refb{enfin} may be expressed as
\be\label{esecond}
S^{(0)}_{BPS} = S_0
+2\pi i J_{3R}\, ,
\ee
which is also not an obvious relation from the point of view of black hole solutions since it
relates the entropy of an extremal black hole to that of a non-extremal black hole.

We shall test both \refb{efirst} and \refb{esecond} for a class of three charge
five dimensional black holes constructed in \cite{Cvetic:1996xz,Cvetic:1996kv}
(see also \cite{Breckenridge:1996sn} for earlier construction for special charges).
The solution has been reviewed in appendix \ref{sa} where we also examine the
consequence of the $\beta\Omega_R=-2\pi i$ equation. Here we just quote some
of the relevant results from that appendix. For BPS black holes we have\cite{9602065} 
\be\label{emassforrep}
M_{BPS} = Q^{(1)} + Q^{(2)} + Q^{(3)}\, ,
\ee
and 
\be\label{etwoSrep}
S^{(0)}_{BPS} = 2\pi \sqrt{4 Q^{(1)}Q^{(2)}Q^{(3)}-J_{3L}^2}\, ,
\ee
where $Q^{(i)}$ for $1\le i\le 3$ are the three U(1) charges carried by the black hole, all
taken to be positive.
Note that the number of gauge fields $n_V$ may still be arbitrary,
but for the particular solution
under consideration, the black hole does not carry charges under these other U(1)
gauge fields.
For the finite temperature solution with $\beta\Omega_R=-2\pi i$, we find in \refb{emassfor},
\be\label{emassforrep2}
M = Q^{(1)} + Q^{(2)} + Q^{(3)}\, . 
\ee
This is the same as \refb{emassforrep}, verifying \refb{efirst}. On the other hand \refb{eoneS}
shows that the entropy of this
black hole is given by
\be\label{eoneSrep}
S_0=2\pi \sqrt{4 Q^{(1)}Q^{(2)}Q^{(3)} - J_{3L}^2} - 2\pi i J_{3R}\, .
\ee
This, together with \refb{etwoSrep},
confirms \refb{esecond}.  Finally, \refb{ebetaexp} shows that 
$J_{3R}$ and the inverse temperature $\beta$ are related as
\be\label{ebeta}
\beta = 2\pi \left(Q^{(1)}Q^{(2)} + Q^{(2)}Q^{(3)}+ Q^{(1)}Q^{(3)} \right)
\left[{1\over \sqrt{4 Q^{(1)}Q^{(2)}Q^{(3)} - J_{3L}^2} } + {i\over J_{3R}}
\right]\, .
\ee
This shows that
$\beta$
remains finite as long as $J_{3R}$ is finite. Furthermore \refb{ebeta}
is consistent with  the scaling relation
\refb{e1.3}.
We recover the extremal black hole  by taking $J_{3R}\to 0$ limit.

\sectiono{Logarithmic correction to the index} \label{slog}

Using \refb{enfin} we can express \refb{ebpsfin} as,
\be\label{elninter}
S_{BPS}=S^{(0)}_{BPS} - n_T \ln L -
\left( {n_T\over 2}  + {n_V\over 2} +{3\over 2}\right) \ln\lambda + \delta\ln Z\, .
\ee
Our goal in this section is to compute $\delta\ln Z$ by computing 
the logarithmic contribution to $\ln Z$ from path integral over various fields
following the procedure described in  \cite{1205.0971}.

As argued in \cite{1205.0971},
only one loop contribution from the massless fields are relevant for
this. Since the metric carries an overall factor of $\lambda^2$, it follows that the eigenvalues
of the kinetic operator for the bosons, involving two derivatives, scale as 
$\lambda^{-2}$ and
the eigenvalues of the kinetic operator for the fermions, involving single derivatives, scale
as $\lambda^{-1}$. Therefore integration over each bosonic mode produces a factor of
$\lambda$ and integration over each fermionic mode produces a factor of $\lambda^{-1/2}$.
The number of modes, although infinite, can be regulated using heat kernel and we get
a finite power of $\lambda$ from integration over these modes. However, in odd dimensions
this power vanishes! Thus it would seem that $\ln Z$ does not receive corrections of order
$\ln\lambda$.

However the analysis described above ignores the presence of zero eigenvalues of the
kinetic operator. If such zero modes are present, their effect needs to be taken into account.
There are two effects: first for each bosonic zero mode we must remove a factor of $\lambda$
from $Z$ and for each fermionic zero mode we must remove a factor of $\lambda^{-1/2}$ from
$Z$ since in the expression based on the heat kernel these modes are treated as if they 
contribute in the same way as the non-zero modes. Second, we need to explicitly evaluate the 
factors of $\lambda$ that may arise from integration over the zero modes. We shall now describe
how this is done for different zero modes.
\begin{enumerate}
\item {\bf Translation zero modes:} The Lorentzian signature black hole in five dimensions will
have four translational zero modes. However for Euclidean signature black hole carrying
angular momenta, the azimuthal angles conjugate to $J_{3L}$ and $J_{3R}$ 
are shifted by $-i\beta\Omega_{L}$ and $2\pi $ respectively as we
go around the Euclidean time circle. Therefore only those translational zero modes that
remain invariant under this twist will be genuine zero modes. The number of such zero
modes will depend on $\Omega_{L}$ (in particular whether or not it vanishes), 
but it is the same number $n_T$ that 
determined the dimension of the momentum vector in our earlier analysis. We shall now
compute the effect of integration over these translation zero modes.

Let $h_{\mu\nu}$ denote the fluctuation mode of the graviton around the Euclidean black
hole solution. We take the path integral measure over $h_{\mu\nu}$ to be $d(\lambda^\alpha
h_{\mu\nu})$ where $\alpha$ is determined by the condition:
\be
\int d(\lambda^\alpha h_{\mu\nu}) \exp\left[- \int d^5 x\sqrt{\det g}\, g^{\mu\rho} g^{\nu\sigma}
h_{\mu\nu} h_{\rho\sigma}
\right]=1\, ,
\ee
$g_{\mu\nu}$ being the background metric. We shall work in the coordinate system in which
$g_{\mu\nu}$'s are given by $\lambda$ independent functions multiplied by $\lambda^2$.
Since $g_{\mu\nu}\sim \lambda^2$, the coefficient of the $h_{\mu\nu} h_{\rho\sigma}$ term
in the
exponent is proportional to $\lambda$. This
determines $\alpha$ to be $1/2$. Therefore the integration measure over $h_{\mu\nu}$ is
$d(\lambda^{1/2} h_{\mu\nu})$.

Now consider the deformation of the metric generated by translation of the solution in some 
particular direction, and label the parameter by $c$. 
Then the zero mode deformation is generated by a diffeomorphism 
$x^\mu\to x^\mu+ c f^\mu(x)$ where $f^\mu(x)$'s will be taken to be
$\lambda$
independent functions which approach some constant vector at infinity lying along
the direction of
translation. 
Thus the
translational zero mode deformations of the metric have the form
\be
h_{\mu\nu} = c \, \left[D_\mu f_\nu(x) + D_\nu f_\mu (x)\right] \sim \lambda^2 \, c\, ,
\ee
where the $\lambda^2$ factor comes from the lowering of the index of $f^\mu$ by the
metric $g_{\mu\nu}$. This gives
\be
d (\lambda^{1/2} h_{\mu\nu}) \sim \lambda^{5/2} dc\, .
\ee
Now, if we confine the black hole inside a box of physical size $L$ as before, then the 
range of $c$ is of order $L/\lambda$ due to the $\lambda^2$ factor in the metric. This
gives the result of a translational zero mode integral to be $\lambda^{5/2}\times L/\lambda \sim
L\lambda^{3/2}$, and we get a net contribution of 
\be
n_T\ln L+ {3\, n_T\over 2} \ln \lambda\, ,
\ee
to $\ln Z$ from $n_T$ translation zero modes. However, as pointed out earlier, 
the heat kernel includes the contribution from each bosonic zero mode to $\ln Z$ 
as $\ln\lambda$. Removing this contribution, we see that the net \emph{extra}
contribution to $\ln Z$ that we have from  $n_T$ translational zero modes is
\be\label{e5.6}
n_T\ln L+ {n_T\over 2} \ln \lambda\, .
\ee

\item Since the black hole carries third components of angular momenta under $SU(2)_L$
and $SU(2)_R$, we can rotate the Lorentzian solution about the 1 and 2 axes of each SU(2)
groups to generate new solutions. This gives four rotational zero modes. However, in the
Euclidean theory the azimuthal angles are twisted by $- i\beta \Omega_L$ and $- i\beta
 \Omega_R=2\pi$
as we go around the time circle, and only those modes that remain invariant under this
twist are genuine zero modes. This eliminates the rotational zero modes of $SU(2)_L$ since
the symmetry generators that generate the zero modes do not commute with $e^{-\beta
\Omega_L J_{3L}}$. However as discussed in \cite{H:2023qko}
in the context of four dimensional black holes,
for the special choice $\beta\Omega_R=-2\pi i$, the twist in the azimuthal angle 
conjugate to $J_{3R}$ becomes
$2\pi$ and now the two rotation generators in $SU(2)_R$ are invariant under this twist.
The contribution from these modes can be analyzed in the same way as for translational modes,
with the only difference that the range of integration over the diffeomorphism parameters 
(analog of $c$ for the translation modes) is a constant instead of order $L/\lambda$.
Therefore compared to the translational zero mode, the rotational zero modes have an
additive contribution of $\ln(\lambda/L)$ to $\ln Z$ per zero mode.
Since \refb{e5.6} gives a contribution of $\ln (L\lambda^{1/2})$ per translation zero mode,
we see that we have
a net extra contribution 
of ${3\over 2}\ln\lambda$ per rotational zero mode. This gives the net \emph{extra}
logarithmic correction to $\ln Z$ from the two rotational zero modes to be:
\be
3\ln\lambda\, .
\ee

\item Since there are $2n$ broken supersymmetries, we also have $2n$ gravitino zero modes.
To find their contribution we proceed as in \cite{H:2023qko}.
If we take the integration measure over the gravitino
fields $\psi_\mu$ to be $d(\lambda^\gamma \psi_\mu)$ then we require
\be
\int d(\lambda^\gamma \psi_\mu) \exp\left[-\int d^5 x\sqrt{\det g} 
g^{\mu\nu} \bar\psi_\mu \psi_\nu
\right]
=1\, .
\ee
Since $g_{\mu\nu}\sim \lambda^2$, the coefficient of the $\bar\psi_\mu \psi_\nu$ term
in the exponent is of order $\lambda^3$. Therefore we must
choose $\gamma=3/2$. On the other hand, the action for the gravitino field takes the form
\be
\int d^5 x\sqrt{\det g}  E_a^\rho
g^{\mu\nu} \bar\psi_\mu \gamma^a \p_\rho \psi_\nu +\cdots
\sim \lambda^2 \int \bar \psi \p \psi \\, ,
\ee
where $E_a^\rho$ is the inverse vierbein. Therefore $J_{3R}$, constructed from the
action using Noether prescription, also carries a factor of $\lambda^2$ and the
part of $J_{3R}$ involving the fermion zero modes $\psi_p$ has the form 
$\lambda^2 c_{pq}\psi_p\psi_q$ for some $\lambda$-independent
constants $c_{pq}$. The integration over the
gravitino zero modes now takes the form
\be\label{egrafin}
\int \prod_{r=1}^{2n} d(\lambda^{3/2}\psi_r) (\lambda^2 c_{pq} 
\psi_p  \psi_q)^n \sim \lambda^{-n}\, .
\ee
We need to multiply this by $\lambda^{1/2}$ for each
of the $2n$ zero modes since the heat kernel counts a factor of $\lambda^{-1/2}$ for each zero
mode. Thus cancels the $\lambda^{-n}$ factor in \refb{egrafin}. Therefore the net 
\emph{extra} logarithmic
correction to $\ln Z$ from the gravitino zero mode integration vanishes.\footnote{Such
cancellations were also present in four dimensions\cite{H:2023qko}. One can in fact
see this cancellation by working with general $D$ where the integration measure over
the gravitino zero mode is $d(\lambda^{(D-2)/2}\psi_\mu)$ and $J_{3R}$ is of
order $\lambda^{D-3}$.}
\end{enumerate}
Combining these results we get the logarithmic correction to $\ln Z$:
\be
\delta\ln Z =  n_T\ln L+  \left( 3 +{n_T\over 2}\right) \ln\lambda\, .
\ee
Substituting this into \refb{elninter} we get
\be
S_{BPS} = S^{(0)}_{BPS}+
\left( {3\over 2} -{n_V\over 2} \right) \ln\lambda\, .
\ee
This is in agreement with the microscopic results as well as the result of near horizon
analysis described in \cite{1109.3706}.

$N_{BPS}=e^{S_{BPS}}$ computed above counts states with fixed value of
$ J_{3L}$ but all $\vec J_{L}^2$. The microscopic analysis also gives this
directly\cite{1109.3706}.
For $J_{3L}=0$,  \cite{1109.3706} considered another quantity that counts all BPS states
with fixed charge and $\vec J_{L}^2=0$, i.e.\ only $SU(2)_L$ singlet states, since the
macroscopic analysis based on near horizon $AdS_2\times S^2$ geometry gave this
result directly. We do not consider it here since in our analysis the ensemble with fixed
$J_{3L}$ but all $\vec J_{L}^2$ is the natural ensemble in which the macroscopic results
are found even for vanishing $J_{3L}$ eigenvalue. However, it is possible to extract from
our result the result for fixed charge and $\vec J_{L}^2=0$ following the analysis described
in \cite{1205.0971,1109.3706}.

\bigskip

\noindent{\bf Acknowledgement}: 
C.C. would like to acknowledge the organizers of the ST4 workshop and IIT Mandi for hospitality during the course of the work.
Research at ICTS-TIFR is supported by the Department of Atomic Energy Government of India, under Project Identification No. RTI4001.
The work of A.S. is supported by ICTS-Infosys Madhava 
Chair Professorship
and the J.~C.~Bose fellowship of the Department of Science and Technology.

\appendix

\sectiono{Classical black holes} \label{sa}

\def\sde{\sinh\delta_e}
\def\sdeo{\sinh\delta_{e1}}
\def\sdet{\sinh\delta_{e2}}
\def\cde{\cosh\delta_e}
\def\cdeo{\cosh\delta_{e1}}
\def\cdet{\cosh\delta_{e2}}
\def\cst{\cos^2\theta}
\def\sst{\sin^2\theta}
\def\ct{\cos\theta}
\def\st{\sin\theta}
\def\non{\nonumber}

\def\sh{\sinh}
\def\ch{\cosh}
\def\de{\delta_{e3}}
\def\deo{\delta_{e1}}
\def\dett{\delta_{e2}}

The goal of this appendix is to verify the relation between BPS and non-BPS entropy and mass
given in \refb{efirst} and \refb{esecond} for classical black hole solutions.

We use the solution given in \cite{Cvetic:1996xz}, labelled by six parameters $m,\de,
\deo,\dett,l_1$ and $l_2$. In $G_N=\pi/8$ units, the metric is given 
by
\ben\label{emetric}
ds^2 &=& \Delta^{1/3} \Bigg[ - {(r^2 + l_1^2 \cos^2\theta+l_2^2\sin^2\theta)
(r^2 + l_1^2 \cos^2\theta+l_2^2\sin^2\theta-2m)\over \Delta} dt^2
\non\\ &+&{r^2 \over (r^2+l_1^2) (r^2+l_2^2)-2mr^2} dr^2 +d\theta^2 \nonumber \\
& +& d\phi d\psi \, {4m\cst\sst\over \Delta} \Bigg\{l_1l_2\Big( (r^2+l_1^2\cst+l_2^2\sst)\non\\ &-&
2m (\sh^2\de \sh^2\deo+\sh^2\de \sh^2\dett+\sh^2\deo \sh^2\dett)\Big)
\non\\ &+& 
2m\Bigg((l_1^2+l_2^2)\ch\deo \ch\dett \ch\de \sh\deo \sh\dett \sh\de 
\nonumber \\ && -2l_1l_2 \sh^2\de\sh^2\deo\sh^2\dett\Bigg) 
\Bigg\} \non\\ &-& {4 m \sst\over \Delta} d\phi dt\, \Bigg\{ (r^2+l_1^2\cst +l_2^2\sst)
(l_1\ch\deo \ch\dett \ch\de \nonumber \\ && \hskip 1.5in - l_2 \sh\deo \sh\dett \sh\de)
%\non\\ &+& 
+ 2ml_2 \sh\deo \sh\dett \sh\de\Bigg\}
\non\\ &-& {4 m \cst\over \Delta} d\psi dt\, \Bigg\{ (r^2+l_1^2\cst +l_2^2\sst)
(l_2\ch\deo \ch\dett \ch\de \nonumber \\ && \hskip 1.5in - l_1 \sh\deo \sh\dett \sh\de)
%\non\\ &+& 
+ 2ml_1 \sh\deo \sh\dett \sh\de\Bigg\} \non\\ &+&
{\sst \over \Delta} d\phi^2 \Bigg\{ (r^2 + 2m\sh^2\de+l_1^2)(r^2 +2m\sh^2\deo +
l_1^2\cst + l_2^2\sst)\non\\ &&(r^2 +2m\sh^2\dett +
l_1^2\cst + l_2^2\sst)  \non\\ &+& 2m\sst \Big( (l_1^2\ch^2\de-l_2^2 \sh^2\de)(r^2+l_1^2\cst
+l_2^2 \sst) \non\\ &+& 4 m l_1 l_2 \ch\deo \ch\dett \ch\de\sh\deo \sh\dett \sh\de
\non\\ &-& 2m\sh^2\deo \sh^2\dett (l_1^2 \ch^2\de+l_2^2 \sh^2\de)-2ml_2^2\sh^2\de
(\sh^2\deo+\sh^2 \dett)\Big)\Bigg\}\non\\ &+&
{\cst \over \Delta} d\psi^2 \Bigg\{ (r^2 + 2m\sh^2\de+l_2^2)(r^2 +2m\sh^2\deo +
l_1^2\cst + l_2^2\sst)\non\\ &&\hskip 1in (r^2 +2m\sh^2\dett +
l_1^2\cst + l_2^2\sst)  \non\\ &+& 2m\cst \Big( (l_2^2\ch^2\de-l_1^2 \sh^2\de)(r^2+l_1^2\cst
+l_2^2 \sst) \non\\ &+& 4 m l_1 l_2 \ch\deo \ch\dett \ch\de\sh\deo \sh\dett \sh\de
\non\\ &-&   2m\sh^2\deo \sh^2\dett (l_2^2 \ch^2\de+l_1^2 \sh^2\de)\nonumber \\
&-& 2ml_1^2\sh^2\de
(\sh^2\deo+\sh^2 \dett)\Big)\Bigg\}
\Bigg]\, ,%\nonumber \\
\een
where
\ben
\Delta &=& (r^2 + 2 m \sh^2 \de + l_1^2 \cos^2\theta + l_2^2 \sin^2\theta)
(r^2 + 2 m \sh^2 \deo + l_1^2 \cos^2\theta + l_2^2 \sin^2\theta)\non\\ &&
\hskip 1in (r^2 + 2 m \sh^2 \dett + l_1^2 \cos^2\theta + l_2^2 \sin^2\theta)\, .
\een
Various other field configurations for this solution can be found in 
\cite{Cvetic:1996xz}, but we shall not
require them for our analysis. 
The mass $2M$, three charges $2Q^{(1)}$, $2Q^{(2)}$, $2Q^{(3)}$ 
and two angular momenta
$J_\phi$, $J_\psi$ are given in terms of these parameters via the 
equations:\footnote{In an earlier version of the paper we mentioned $G_N$ to be $\pi/4$, but
the correct value that is compatible with the expressions for the angular momenta
given in \refb{emasses} and the entropy given in \refb{eentropy}
is $\pi/8$. This however changes the masses and charges
to $2M$ and $2Q^{(i)}$ with $M$ and $Q^{(i)}$ given in \refb{emasses}.
We thank Kanhu Kishore Nanda and Amitabh Virmani for pointing this out.
}
\ben \label{emasses}
&& Q^{(1)}=2\, m\, \ch\deo \sh\deo, \qquad Q^{(2)} = 2\, m\, \ch\dett \sh\dett, \qquad
Q^{(3)}= 2\, m\, \ch\de \sh\de, \non\\
&& M = \sqrt{m^2+(Q^{(1)})^2}+ \sqrt{m^2+(Q^{(2)})^2}+ \sqrt{m^2+(Q^{(3)})^2}, \non\\
&& J_\phi = 4m (l_1\ch\deo\ch\dett\ch \de - l_2\sh\deo\sh\dett\sh \de), \non\\ &&
J_\psi = 4m (l_2\ch\deo\ch\dett\ch \de - l_1\sh\deo\sh\dett\sh \de)\, .
\een
This gives
\ben \label{ea4}
J_{3L} = {1\over 2} (J_\phi-J_\psi) = 2m(l_1-l_2) 
(\ch\deo\ch\dett\ch \de + \sh\deo\sh\dett\sh \de)\, ,
\non\\
J_{3R} = {1\over 2} (J_\phi+J_\psi) = 2m(l_1+l_2) 
(\ch\deo\ch\dett\ch \de - \sh\deo\sh\dett\sh \de)\, . %\nonumber \\
\een

The locations $r_\pm$ of the outer and inner horizons are at $r=r_\pm$ where
\be
r_\pm ^2= m -{1\over 2}l_1^2 -{1\over 2}l_2^2 \pm {1\over 2} \sqrt{(l_1^2-l_2^2)^2 + 4m
(m-l_1^2-l_2^2)}\, .
\ee
The entropy is given by\cite{Cvetic:1996kv}
\ben\label{eentropy}
S_0 &=& 4\pi \Bigg[ m \left\{ 2m - (l_1-l_2)^2 \right\}^{1/2} \left\{ \ch\de \ch\deo\ch\dett
+ \sh\de \sh\deo\sh\dett\right\} \nonumber \\
&+& m \left\{ 2m - (l_1+l_2)^2 \right\}^{1/2} \left\{ \ch\de \ch\deo\ch\dett
- \sh\de \sh\deo\sh\dett\right\}\Bigg] \nonumber \\
&=& 4\pi \Bigg[ \left\{ 2 m^3 (\ch\de \ch\deo\ch\dett
+ \sh\de \sh\deo\sh\dett)^2 -{1\over 4} J_{3L}^2\right\}^{1/2} \nonumber \\
&+& \left\{ 2 m^3 (\ch\de \ch\deo\ch\dett
- \sh\de \sh\deo\sh\dett)^2 -{1\over 4} J_{3R}^2\right\}^{1/2}\Bigg] \, .
%\nonumber \\
\een
Now it follows from \refb{emasses} that $M$, $Q^{(1)}$, $Q^{(2)}$ and $Q^{(3)}$ are functions of
$m$, $\delta_{e1}$, $\delta_{e2}$ and $\de$, and inverting these relations
$m$, $\de$, $\delta_{e1}$ and $\delta_{e2}$ can be expressed as functions of
$M$, $Q^{(1)}$, $Q^{(2)}$ and $Q^{(3)}$. However
these relations do not involve $J_{3L}$
and $J_{3R}$. Therefore we get,
\be
\beta\Omega_R = {\p S_0\over \p J_{3R}}
=-\pi J_{3R} \left\{ 2 m^3 (\ch\de \ch\deo\ch\dett
- \sh\de \sh\deo\sh\dett)^2 - {1\over 4} J_{3R}^2\right\}^{-1/2}\, ,
\ee
\be
\beta\Omega_L = {\p S_0\over \p J_{3L}}
=-\pi J_{3L} \left\{ 2 m^3 (\ch\de \ch\deo\ch\dett
+ \sh\de \sh\deo\sh\dett)^2 -{1\over 4} J_{3L}^2\right\}^{-1/2}\, .
\ee

For studying the index, we want to set $\beta\Omega_R$ to $-2\pi i$. 
If we use the convention
\be \label{econv}
(-J_{3R}^2)^{1/2} = - i\, J_{3R} \qquad  \Leftrightarrow \qquad (-J_{3R}^2)^{-1/2} =i /J_{3R}\, ,
\ee
then we get
\be
m=0, \qquad J_{3R} \quad  \hbox{finite}\, .
\ee
Naively for $m=0$ the charges, mass and 
angular momenta given in \refb{emasses} vanishes. However we take the $m\to 0$ keeping the
charges and angular momenta fixed. This give
\ben
&& \delta_{ei} =  {1\over 2} \sinh^{-1} { Q^{(i)}\over m}, 
\qquad \hbox{for $1\le i\le 3$}\, , \nonumber \\ &&
l_1-l_2= {J_{3L} \over 2\, m
(\ch\deo\ch\dett\ch \de + \sh\deo\sh\dett\sh \de)}\, ,
\non\\ &&
l_1+l_2  = {J_{3R}\over  2m
(\ch\deo\ch\dett\ch \de - \sh\deo\sh\dett\sh \de)}\, .
\een
We shall take the $Q^{(i)}$'s to be positive.
It is easy to check that in the $m\to 0$ limit,
\ben \label{ea12}
l_1-l_2 &\simeq& { J_{3L} \, \sqrt{m} \over \sqrt{2 Q^{(1)}Q^{(2)} Q^{(3)}}},
\nonumber \\
l_1+l_2 &=& {\sqrt 2 J_{3R} \sqrt{Q^{(1)}Q^{(2)}Q^{(3)}}\over \sqrt m \left(
Q^{(1)}Q^{(2)} + Q^{(2)}Q^{(3)}+ Q^{(1)}Q^{(3)}\right)}
\, .
\een
Therefore in this limit, $l_1-l_2$ vanishes but $l_1+l_2$ diverges.

We can compare this with the extremal limit studies in \cite{Cvetic:1996xz}. It is essentially the
same limit, except that there both $l_1$ and $l_2$ were taken to vanish as we take the
$m\to 0$ limit. Therefore in this limit $J_{3R}$ vanishes. We can recover the extremal
limit from our formula by taking the $J_{3R}\to 0$ limit in our final formula.

We now note from \refb{emasses} that in the $m\to 0$ limit, we have
\be\label{emassfor}
M = Q^{(1)} + Q^{(2)} + Q^{(3)}\, .
\ee
This is the same as the BPS mass formula found in \cite{Cvetic:1996xz} and confirms \refb{efirst}.
In fact, the agreement with the BPS formula is a simple consequence of the fact that
\refb{emassfor} is independent of $J_{3R}$ and hence the formula remains the same
in the $J_{3R}\to 0$ limit. 

In the $m\to 0$ limit with fixed charges and angular momenta, the entropy 
\refb{eentropy} takes the form:
\be\label{eoneS}
S_0=2\pi \sqrt{4 Q^{(1)}Q^{(2)}Q^{(3)} - J_{3L}^2} - 2\pi i J_{3R}\, ,
\ee
where we used the convention \refb{econv} as before. 
On the other hand the classical entropy of a supersymmetric black hole is given by
\be\label{etwoS}
S^{(0)}_{BPS} = 2\pi \sqrt{4 Q^{(1)}Q^{(2)}Q^{(3)}-J_{3L}^2}\, .
\ee
\refb{eoneS} and \refb{etwoS} confirm \refb{esecond}. \refb{eoneS} also shows that in order to
get a real $S_0>S^{(0)}_{BPS}$, as would be the case if we had a regular event horizon, 
$J_{3R}$ should lie along the positive
imaginary axis. \refb{ea12} now shows that $(l_1+l_2)$ should be imaginary.

We can also compute the inverse temperature $\beta$ in this limit. The result is
\be\label{ebetaexp}
\beta = 2\pi \left(Q^{(1)}Q^{(2)} + Q^{(2)}Q^{(3)}+ Q^{(1)}Q^{(3)} \right)
\left[{1\over \sqrt{4 Q^{(1)}Q^{(2)}Q^{(3)} - J_{3L}^2} } +{i\over J_{3R}}
\right]\, .
\ee
This remains finite, showing that we have a non-extremal black hole solution in this
limit. We recover the extremal black hole  by taking $J_{3R}\to 0$ limit.

\def\q{\theta}

We can also write down an expression for the metric in this limit. For this, let us define a
new radial variable $\rho$ via the equation
\be
\rho^2 \equiv r^2   - {1\over 2} (r_+^2+r_-^2)\, .
\ee
We also define
\ben
&& Q_P \equiv Q^{(1)} Q^{(2)} + Q^{(2)} Q^{(3)} + Q^{(3)} Q^{(1)}, \qquad
Q_S \equiv Q^{(1)} + Q^{(2)}  + Q^{(3)} , \nonumber \\
&& Q_T \equiv Q^{(1)}Q^{(2)} Q^{(3)}, \qquad Q_R \equiv  (Q^{(1)})^2 (Q^{(2)})^2 + (Q^{(2)})^2
 (Q^{(3)})^2 + (Q^{(3)})^2 (Q^{(1)})^2\, .\nonumber \\
\een
Then in the limit $m\to 0$ with the charges and angular momenta fixed, 
we get,
\be
\Delta = \frac{1}{8 (Q_P)^3} \prod_{i=1}^3\Big[ \cos(2\q) J_{3L} J_{3R} + 2 \left(Q^{(i)} + \rho^2\right) Q_P \Big] \, ,
\ee
and
the metric
takes the form:
\ben\label{emetriclimit}
ds^2 &=& \Delta^{1/3} \Bigg[ - \frac{\left\{\cos (2 \theta ) J_{3L} J_{3R}+2 \rho^2 Q_P\right\}^2}
{4 \, \Delta  \, Q_P^2} dt^2 +\frac{4Q_P^2 \rho^2}{4 \, Q_P^2\, \rho ^4 -J_{3R}^2 \left(J_{3L}^2-4 Q_T\right)} d\rho^2
+d\theta^2 \nonumber \\
&-& d\phi dt \, \frac{\sin ^2\theta  \left\{2 \rho ^2 Q_P \left(J_{3L}+J_{3R}\right)+\cos (2 \theta ) J_{3L} J_{3R} \left(J_{3L}+J_{3R}\right)+4 Q_T J_{3R}\right\}}{2\, \Delta \, Q_P}
\non\\
&-& d\psi dt\, \frac{\cos ^2\theta  \left\{2 \rho ^2 Q_P \left(J_{3R}-J_{3L}\right)+\cos (2 \theta ) J_{3L} J_{3R} \left(J_{3R}-J_{3L}\right)+4 Q_T  J_{3R}\right\}}{2\,  \Delta\,  Q_P}
\non\\ & +&  d\phi d\psi \, \frac{\sin ^2\theta  \cos ^2\theta  }{2\, Q_P^3 \Delta}\times\Bigg[ Q_P 
\Bigl\{ J_{3L}^2 Q_P^2-J_{3R}^2 Q_R
 +2 J_{3R}^2 Q_T  \left(2 \rho ^2 + Q_S\right)\Bigr\} \nonumber \\ && \hskip 2in 
  +2 Q_T  \cos (2 \theta ) J_{3L} J_{3R}^3 \Bigg]
\non\\ 
&+&
{\sst \over \Delta} d\phi^2 \Bigg[
\rho^6 + \rho^4 \Bigg\{ \frac{(4 \cos^2 \q - 1) J_{3L} J_{3R}}{2 Q_P}+Q_S\Bigg\} \nonumber \\
&+& \frac{\rho^2}{4 Q_P^2} \Bigg\{ 4 \cos ^2\theta  J_{3L} J_{3R} Q_P Q_S+ \left(4 \cos ^4\theta -1\right) J_{3L}^2 J_{3R}^2+4 Q_T \sin ^2\theta  J_{3R}^2+4 Q_P^3 \Bigg\}
\nonumber\\
&+& \frac{1}{8 Q_P^3} \Bigg\{-2 \sin ^2\theta  Q_P^3 \left(J_{3R}^2+J_{3L}^2\right)+2 J_{3R}^2 Q_P Q_S \left(\cos (2 \theta ) J_{3L}^2 + 4 Q_T \sin ^2\theta \right)
\nonumber\\
&&\quad +4 J_{3L} J_{3R} \Big(Q_T \sin ^2\theta  \cos (2 \theta ) J_{3R}^2+Q_P^3
\cos ^2\theta  \Big)+8 Q_T  Q_P^3 +\cos ^2(2 \theta ) J_{3L}^3 J_{3R}^3 \Bigg\} \Bigg]
\nonumber \\ &+&
{\cst \over \Delta} d\psi^2 
\Bigg[
\rho^6 + \rho^4 \Bigg\{ \frac{(1 - 4 \sin^2\q) J_{3L} J_{3R}}{2 Q_P}+Q_S\Bigg\} \non \\
&+& \frac{\rho^2}{4 Q_P^2} \Bigg\{-4 \sin ^2\theta  J_{3L} J_{3R} Q_P Q_S+(4 \sin^4\theta-1 ) J_{3L}^2 J_{3R}^2+4 Q_T \cos ^2\theta  J_{3R}^2+4 Q_P^3 \Bigg\} \non\\
&-&\frac{1}{8Q_P^3}  \Bigg\{ 2 \cos ^2\theta  Q_P^3 \left(J_{3L}^2+J_{3R}^2\right)+2 J_{3R}^2 Q_P Q_S \left(\cos (2 \theta ) J_{3L}^2-4 Q_T \cos ^2\theta \right)   \non \\
&+&4 J_{3L} J_{3R} \Big(Q_P^3\sin ^2\theta  -Q_T \cos ^2\theta  \cos (2 \theta ) J_{3R}^2\Big)+J_{3L}^3 J_{3R}^3 \cos^{2}(2\theta) -8 Q_T Q_P^3\Bigg\}
\Bigg]\, .
\een
From this we see that the metric expressed in the $\rho,t,\theta,\phi,\psi$ coordinate
system is finite in the $m\to 0$ limit. We should note however that in this limit $(r_+^2+r_-^2)$
diverges and hence $\rho^2$ and $r^2$ are related by an infinite shift. Therefore the metric
expressed in the $r,t,\theta,\phi,\psi$ coordinate
system would not be finite. In particular the horizon at $r=r_+$ will be at infinite value of $r$.

Even though we have written down the Lorentzian version of the black hole metric, it is the
Euclidean version of this solution that provides the saddle point for the gravitational
path integral for the index.
It should be possible to check that this metric (and other fields whose form can be
found in \cite{Cvetic:1996xz}) 
admit
Killing spinors and therefore describe a supersymmetric configuration. We have not checked
this. However the fact that the solution saturates the BPS mass bound is a strong indication
that this is indeed the case.

A Mathematica notebook containing various computations in this appendix is being submitted
to the arXiv with this paper.

\end{document}